\documentclass[aps,physrev,reprint,superscriptaddress]{revtex4-2}

\usepackage{chemformula}
\setchemformula{kroeger-vink}
\usepackage{siunitx}
\DeclareSIUnit\angstrom{\text{Å}}
\DeclareSIUnit\atmosphere{\text{atm}}
\usepackage[inline]{enumitem}
\usepackage{physics}
\usepackage{graphicx}
\usepackage{mathtools}

\begin{document}
\newcommand{\red}[1]{\textcolor{red}{#1}}
\newcommand{\blue}[1]{\textcolor{blue}{#1}}

\title{Density Functional Theory Study of Th-doped LiCAF and LiSAF for Nuclear Clock Applications}


\author{Martin Pimon}
\affiliation{Institute for Theoretical Physics, TU Wien, Wiedner Hauptstrasse 8-10/E136, 1040 Vienna, Austria}

\author{Tobias Kirschbaum}
\affiliation{University of Würzburg, Institute of Theoretical Physics and Astrophysics, Am Hubland, 97074 Würzburg, Germany}

\author{Thorsten Schumm}
\affiliation{Institute for Atomic and Subatomic Physics, TU Wien, Stadionallee 2, 1020 Vienna, Austria}

\author{Adriana P\'alffy}
\affiliation{University of Würzburg, Institute of Theoretical Physics and Astrophysics, Am Hubland, 97074 Würzburg, Germany}

\author{Andreas Grüneis}
\affiliation{Institute for Theoretical Physics, TU Wien, Wiedner Hauptstrasse 8-10/E136, 1040 Vienna, Austria}


\date{\today}

\begin{abstract}
Thorium-doped \ch{LiCaAlF_6} and \ch{LiSrAlF_6} (\ch{Th:LiCAF} and \ch{Th:LiSAF}) are promising crystals for a solid-state nuclear clock based on the 8 eV transition in $^{229}$Th; however, their complex crystal structures complicate understanding the atomic arrangement of the thorium defects. In this work, density functional theory simulations are employed to systematically investigate these systems, including temperature-dependent effects and environmental conditions of fluorine saturation and deficiency. We investigated 20 distinct charge compensation schemes for each material, revealing lower defect formation energies in \ch{Th:LiSAF} than in \ch{Th:LiCAF}. This suggests that the former may attain a higher concentration of thorium nuclei. Furthermore, we calculated the electric field gradient for the lowest energy structure per compensation pathway. Our investigation shows that results previously reported in the literature apply only to a subset of experimental conditions.
\end{abstract}


\maketitle

\section{Introduction}
The thorium-229 isotope has an exceptionally low-energy first nuclear excited state~\cite{NuclearIsomersMahesh2024}. Recently, Zhang et al.~\cite{FrequencyRatioZhang2024} performed high-precision laser spectroscopy of $^{229}$Th nuclei doped in \ch{CaF_2} and reported a transition energy of \SI{8.355732812685 \pm 0.000000000008}{\electronvolt} above the ground state. Due to its narrow line width, this transition is a candidate for a new frequency reference standard~\cite{The229ThoriumThirol2019, The229TVonDe2020, TheThoriumIsoThirol2024} and its practical implementation is commonly referred to as \emph{nuclear clock}~\cite{NuclearLaserSPeik2003, SingleIonNuclCampbe2012, TheThorium229Beeks2021}. Its unprecedented sensitivity to temporal variations of fundamental constants and dark matter candidates is predicted to enable exploration of physics beyond the Standard Model~\cite{TestingTimeVaBereng2010, TheSearchForSafron2019, SensitivityOfFadeev2020, NuclearClocksPeik2021, OnTheSensitivCaputo2024}.

Attempts to directly excite the nucleus in neutral atoms by laser irradiation would inadvertently lead to ionization since the first ionization potential of Th is \SI{6.3}{\electronvolt} and thus lower than the transition to the nuclear isomer~\cite{SpectroscopyOfGoncha2006}. One way to remove outer shell electrons and simultaneously increase the number of Th nuclei in the target is to form an ionic solid with thorium and other elements. Because the scarcity of the isotope makes single-crystal growth impractical~\cite{TheThorium229Beeks2021}, doping it into bulk insulators that are transparent to the isomeric transition wavelength is a more viable solution~\cite{PerformanceOfKazako2012}. For this purpose, several large-band gap host crystals have been proposed~\cite{OpticalSpectroHehlen2013}, and a few have been successfully grown and tested experimentally: \ch{Th:CaF_2}, \ch{Th:MgF_2}, and \ch{Th:LiSrAlF_6} (\ch{Th:LiSAF})~\cite{FrequencyRatioZhang2024, LaserExcitatioTiedau2024, LaserExcitatioElwell2024, RadiativeDecayPineda2025}.

Numerical simulations are essential for gaining insights into the microscopic structure of defects in such crystals, enabling calculation of nuclear sub-level splitting caused by an interaction with the Electric Field Gradient (EFG) at the thorium position and the nuclear quadrupole moment. Additionally, calculating the defect formation energy for various structures provides the thermodynamic probability of these structures occurring during crystal growth.

For various defect configurations of \ch{Th:CaF_2} and \ch{Th:MgF_2}, EFG values and defect formation energies were calculated using Density Functional Theory (DFT)~\cite{Sup229SupTDessov2014, DftCalculationPimon2020, AbInitioStudyPimon}. The number of viable defect configurations in these binary hosts is limited compared to the quaternary hosts \ch{LiSAF} and \ch{LiCAF}, making their defect structures more difficult to predict. \textcite{ComputerModellJackso2009} performed the initial investigation into these systems using a classical potential. \textcite{BandGapCalculPimon2022} simulated \ch{Th:LiCAF} using DFT, providing defect formation energies, evaluated using a canonical ensemble. Another DFT study of \ch{Th:LiSAF} based on the data generated in Ref.~\cite{BandGapCalculPimon2022} reported the lowest energy structure, but did not publish its defect formation energy~\cite{LaserExcitatioElwell2024}.

With the present study, we aim to improve previous research on \ch{Th:LiSAF} and \ch{Th:LiCAF} in three key ways: Firstly, we aim at an unbiased approach by generating a large set of structural configurations for the \ch{Th:LiSAF} system. Secondly, we report defect formation energies of the various charge compensation schemes in \ch{Th:LiSAF} and go beyond those reported for \ch{Th:LiCAF}~\cite{BandGapCalculPimon2022} by considering tunable chemical potentials in the grand canonical ensemble and including temperature dependence. Thirdly, we present the electric field gradients for the lowest energy structures per charge compensation scheme for both materials.

\section{Methods}

We follow the approach outlined in Freysoldt et al.~\cite{FirstPrincipleFreyso2014} to calculate defect formation energies. A breakdown of the complete workflow in this study is as follows:
\begin{enumerate*}[label=\Alph*.]
 \item We compile a list of compounds in the Li-Ca(Sr)-Al-F-Th system.
 \item We calculate the chemical potentials of these elements.
 \item We generate a list of potential charge compensation schemes in \ch{Th:LiSAF}, generate defect structures for each compensation scheme, and calculate their electronic internal energies. Together with the dataset from Ref.~\cite{BandGapCalculPimon2022}, we only select the defect structures with the lowest energy per compensation scheme and calculate the vibrational contribution to their free energy.
 \item We simulate a lower doping concentration for each structure to obtain a correction in the dilute doping limit and to calculate the electric field gradient on thorium.
 \item We determine the defect formation energy for each structure under different experimental conditions.
\end{enumerate*}

\subsection{Compound Selection}
We select compounds comprising combinations of cations Li, Ca, Sr, Al, and Th with anions F that have been experimentally observed and characterized. We gather structural data for the solid phases by querying the Materials Project database~\cite{TheMaterialsAOngS2015, CommentaryTheJain2013, PythonMaterialOngS2013} and cross-validate the results with the Inorganic Crystal Structure Database~\cite{RecentDevelopmZagora2019}. Due to the impracticality of an exhaustive analysis encompassing all possible combinations, we restrict our search to fluorides because of their similarity to thorium-doped LiCAF and LiSAF. In cases of phase transitions, we utilize the stable phase determined by DFT simulations at \SI{0}{\kelvin}. We further include metastable compounds slightly above the convex hull~\cite{A_critical_exam_Bartel_2020} to restrict the chemical potential domain, and we also include metallic elementary phases to define the upper bound of their element's chemical potential (see Sec.~\ref{sec: Calc Mu}).

We now list all compounds considered in this study. The single-component metals \ch{Li}, \ch{Ca}, \ch{Sr}, \ch{Al}, and \ch{Th}; the compounds \ch{LiF}, \ch{CaF_2} (\ch{SrF_2}), and \ch{AlF_3} are the constituents of the host crystal \ch{LiCAF} (\ch{LiSAF}), along with \ch{ThF_4} to create the defect. \ch{Li_3AlF_6}, \ch{LiThF_5}, and \ch{Li_3ThF_7} contain neither Ca nor Sr. \ch{CaAlF_5}, and \ch{Ca_2AlF_7} contain Ca but not Sr, whereas \ch{SrAlF_5} and \ch{Sr_5Al_2F_{16}} contain Sr but not Ca. Finally, \ch{F_2} is the only gaseous phase we include.

\subsection{Calculation of Chemical Potentials}
\label{sec: Calc Mu}
The crystals are typically grown at ambient pressure and melting temperature of the host materials (\SI{1082}{\kelvin} for LiCAF and \SI{1038}{\kelvin} for LiSAF) and cooled slowly~\cite{Klimm2000, GrowthAndCharBeeks2023}. Thus, the appropriate thermodynamic potential to describe the process is the Gibbs free energy $G(T) = \mu(T) N = F(T) + pV(T)$, where $T$ represents absolute temperature, $\mu$ is the chemical potential, $N$ is the number of formula units per unit cell, $F$ is the Helmholtz free energy, $p = \SI{1}{\atmosphere}$ is pressure, and $V$ is the unit cell volume.

When simulating solid phases, we calculate the Helmholtz free energy $F(T) = U_{\text{el}} + U_{\text{vib}} - T S_{\text{vib}} (T)$, where we obtain the electronic internal energy $U_{\text{el}}$ from DFT calculations, and the vibrational zero point energy $U_{\text{vib}}$, together with the vibrational entropy $S_{\text{vib}}$ from a frozen phonon simulation~\cite{phonopy1, phonopy2}. We neglect the electronic contribution to the entropy, which is only relevant for metallic phases in the temperature range of this study. Also note that the impact of thermal volume expansion work on the total energy is negligible, and we disregard it by fixing the cell volumes to the DFT equilibrium values $pV(T) = pV_{\text{DFT}}$.

For the gaseous \ch{F_2} phase, we calculate the entropy as $S = S_{\text{trans}} + S_{\text{vib}} + S_{\text{rot}}$. We use the Sackur-Tetrode equation for the translational part $S_{\text{trans}}$, the entropy of the quantum harmonic oscillator for the vibrational part $S_{\text{vib}}$, and the entropy of the quantum rigid rotor for the rotational part $S_{\text{rot}}$ (see also Appendix~\ref{sec:chempot}). To this end, we calculate the fundamental vibrational frequency and the equilibrium bond length with DFT, and use $m_{\ch{F}} = \SI{18.998}{\dalton}$~\cite{StandardAtomicProhas2022}. In addition, we include volume expansion work by applying the ideal gas law $pV(T) = N k_B T$.

We calculate the chemical potentials of the elements Li, Sr, Ca, Al, F, and Th by solving a system of inequalities
\begin{equation}
    \begin{aligned}
        \mu_{\ch{Li}} + \mu_{\ch{Ca}} + \mu_{\ch{Al}} + 6\mu_{F} &= \mu_{\ch{LiCAF}},\\
        \mu_{\ch{Th}} + 4 \mu_{\ch{F}} &= \mu_{\ch{ThF_4}}, \\
        \mu_{\ch{Li}} &\leq \mu_{\ch{Li}}[\text{metal}],\\
        &\vdotswithin{=} \\
        \mu_{\ch{F}}  &\leq \mu_{\ch{F_2}}[\text{gas}]/2,\\
        \mu_{\ch{Li}} + \mu_{\ch{F}} &\leq \mu_{\ch{LiF}},\\
        &\vdotswithin{=} \\
    \end{aligned}
\end{equation}
The chemical potential of a compound cannot exceed that of its allotrope; the first equality relation thus confines the chemical environment to that of the host crystal. The second equality fixes the chemical potential of thorium. For the same reason, the chemical potential of each species cannot be larger than its elementary allotrope's, but it can always be smaller. The remaining inequalities provide information about the chemical potential of each species within various compounds. Consequently, these relations may restrict the codomain for the chemical potential of a species. We then employ a numerical routine that finds the solution for all $\mu_i$ under two conditions: maximizing and minimizing $\mu_{\ch{F}}$, corresponding to fluoride saturation and deficiency. We chose these conditions because crystal growth was performed in a fluorinated atmosphere in thorium-doped calcium fluoride to enhance transmission properties~\cite{OpticalTransmiBeeks2024}.

\subsection{Charge Compensation Schemes}

In examining potential charge compensation mechanisms and initial defect geometries, we adopt an approach analogous to that outlined in Ref.~\cite{BandGapCalculPimon2022}. In essence, our strategy involves imposing a constraint on the number of atoms of a particular species $n_i$, with $i \in \lbrace \text{Li, Ca, Sr, Al, F, Th} \rbrace$, that are introduced ($n_i > 0$) and removed ($n_i < 0$) from the crystal lattice to form the defect, with a limit of five such atoms ($\sum_i \vert {n_i}\vert = 5$).

Intending to generate supercells with lattice vectors of comparable magnitude, we use $2 \times 2 \times 1$ primitive unit cells of the host crystals, resulting in 80 host atoms. When thorium substitutes an atom in the cell, it results in an imbalance between its number of valence electrons ($=4$) and that of the substituted atom. We assume the defect is closed-shell and fully compensated by vacancies or interstitials, constrained to be near the thorium defect. We initially place interstitial atoms on high-symmetry Wyckoff positions. The maximal distance parameter controls the number of atomic arrangements. We aim to generate at least 20 symmetrically unique structures for each of the 20 assessed charge compensation mechanisms. In total, we generate 1812 structures for \ch{Th:LiSAF} and use 1567 structures for \ch{Th:LiCAF}~\cite{BandGapCalculPimon2022}.

We then perform relaxations on these supercells. When interatomic forces are smaller than \SI{10}{\milli \electronvolt \per \angstrom}, we compare the internal energies associated with each charge compensation scheme and select the structure corresponding to the lowest energy state. Subsequently, we calculate the vibrational Helmholtz free energy for the selected structures, incorporating a non-analytical term correction for insulators derived from the Born effective charges and the static dielectric tensor~\cite{nac1, nac2, nac3}.

\subsection{Dilute Doping Limit}

In practice, the defect concentration is low enough to simplify the system to the dilute doping limit. To address this, we correct the electronic internal energy of the $2 \times 2 \times 1$ supercells by simulating a larger supercell with $4 \times 4 \times 2$ repetitions of the primitive unit cell (640 host atoms). After inserting the smaller (doped, relaxed) structure into the larger (pristine) supercell, we perform another relaxation of the system. Then, we apply a correction
to the electronic internal energy $E_{\text{corr}} = U_{\text{el}}[X_{4\times4\times2}] - U_{\text{el}}[X_{2\times2\times1}] - 7U_{\text{el}}[\text{bulk}_{2\times2\times1}]$, where $X$ denotes the supercell containing the thorium defect and ``bulk'' the pristine host crystal. In the larger supercell, we also compute the electric field gradient $V_{ij} = \pdv{V}{x_i}{x_j}$ at the site occupied by thorium.

\subsection{Defect Formation Energies}
Under the conditions and approximations described above, the defect formation energy is~\cite{FirstPrincipleFreyso2014}
\begin{equation}
    G^f[X](T) = F[X](T) - F[\text{bulk}](T) - \sum_i n_i \mu_i (T) + E_{\text{corr}},
\end{equation}
where $G^f$ is the Gibbs free energy of formation, $F$ is the Helmholtz free energy including the electronic internal energy and vibrational contributions, $n_i$ is the number of atoms removed or added to the supercell $X$ to form the defect, $\mu_i$ is the chemical potential for the respective element $i$, and $E_{\text{corr}}$ is the correction term to account for the dilute doping limit.

\subsection{Computational Details}
We employ the Vienna Ab-initio Simulation Package (VASP)~\cite{vasp1, vasp2, vasp3, vasp4, vasp5} version \texttt{6.4.3} to compute internal electronic energies and interatomic forces using the approximation to the exchange-correlation potential by \textcite{GeneralizedGraPerdew1996}~(PBE). We utilize the pseudopotentials \texttt{Li\_sv\_GW}, \texttt{Ca\_sv\_GW}, \texttt{Sr\_sv\_GW}, \texttt{Al\_GW}, \texttt{Th}, and \texttt{F\_h\_GW}.

For pristine crystals, we implement a convergence study by adjusting both the $k$-mesh, energy cutoff, and cell volume until they reach a convergence threshold of $\Delta U_{\text{el}} \leq \SI{1}{\milli \electronvolt \per atom}$ compared to the most accurate calculation. We evaluate doped supercells only at the $\Gamma$-point in $k$-space. When calculating Born effective charges and the static dielectric tensor for doped and pristine crystals, we increase the k-space sampling by doubling the number of points in each direction. We conducted the calculations of \ch{Th:LiCAF} and \ch{Th:LiSAF} with the plane wave basis set having a cutoff of 950 eV.

We calculated vibrational properties using phonopy~\cite{phonopy1, phonopy2} version \texttt{2.21.2} in the frozen phonon approach in the harmonic approximation, targeting supercells with lattice vectors of approximately equal length and containing at least around 100 atoms.

We implemented many parts of the workflow using AiiDA~\cite{aiida1, aiida2, aiida3} (aiida-core version \texttt{2.5.1} with the plugins aiida-vasp~\cite{aiida-vasp} (version \texttt{3.1.0}) and aiida-phonopy~\cite{aiida-phonopy1, aiida-phonopy2} (version \texttt{1.1.4}).

\section{Results}

To briefly describe the stoichiometry for each compensation scheme, we use a notation for the numbers $n_i$: The order of the elements $i$ is Li, Ca (Sr), Al, F, and Th; a bar indicates a negative sign. To give an example, $0\bar{2}001$ corresponds to the removal of two Ca (Sr) atoms, and the addition of one Th atom ($\ch{Th_{Ca}^{..}} + \ch{v_{Ca}^{''}}$ in Kröger-Vink notation~\cite{RelationsBetweKroger1956}). We use the shorthand notation when a concise representation is advantageous, and Kröger-Vink notation when conveying more detailed information about the defect structure.

\begin{figure*}
    \includegraphics{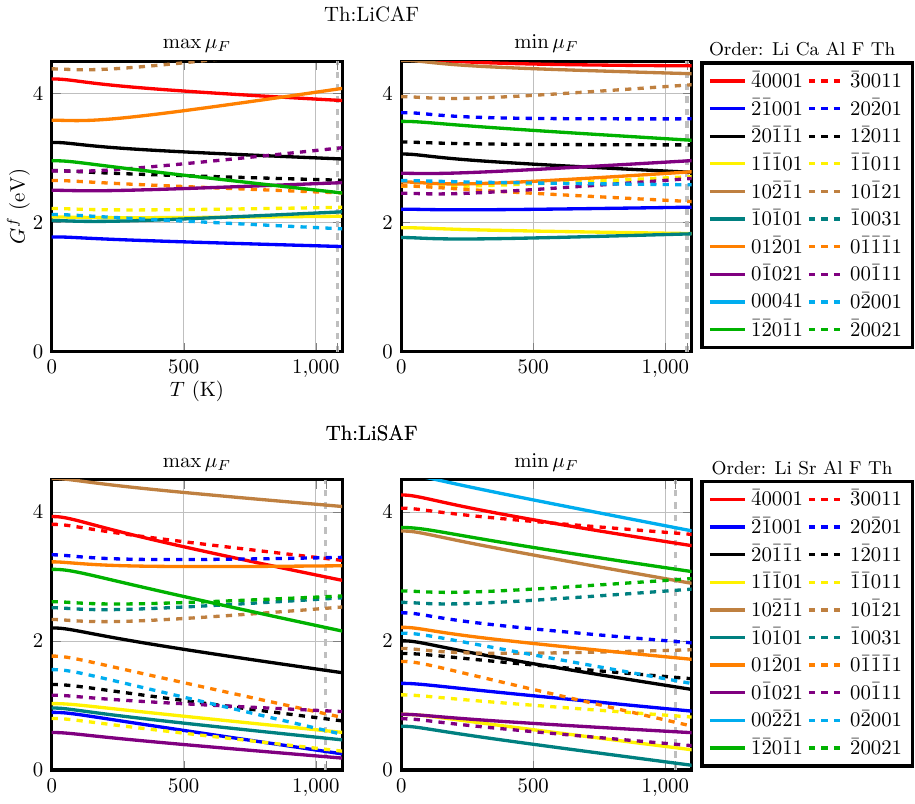}
    \caption{\label{fig:dfe} Defect formation energy of the lowest energy structure per compensation scheme in \ch{Th:LiCAF} (top row) and \ch{Th:LiSAF} (bottom row). A maximum threshold of \SI{4.5}{\electronvolt} is imposed for improved visibility of the more relevant compensation schemes. Dashed vertical lines indicate the host crystal's melting point. Left: $\mu_{\ch{F}}$ maximized. Right: $\mu_{\ch{F}}$ minimized.}
\end{figure*}

Fig.~\ref{fig:dfe} shows the temperature-dependent defect formation energies for the lowest energy structure per charge compensation scheme for \ch{Th:LiCAF} and \ch{Th:LiSAF}. See also the Appendix for a breakdown of different energy contributions. Table~\ref{tab:EFG} presents the calculated and diagonalized EFG values, following the standard convention: $\abs{V_{zz}} \geq \abs{V_{yy}} \geq \abs{V_{xx}}$ and $\eta = \frac{V_{xx} - V_{yy}}{V_{zz}}$.


%


%

\begin{table}
\caption{\label{tab:EFG} $V_{zz}$ and $\eta$ components of the electric field gradient at the thorium site of the lowest energy structure per charge compensation scheme.}
\begin{ruledtabular}
    \sisetup{round-mode=places, round-precision=2}
    \begin{tabular}{c r c r c}
    & \multicolumn{2}{c}{\ch{Th:LiCAF}} & \multicolumn{2}{c}{\ch{Th:LiSAF}} \\
    Comp. & $V_{zz}$ (\si{\volt \per \angstrom^2}) & $\eta$ & $V_{zz}$ (\si{\volt \per \angstrom^2}) & $\eta$ \\  \hline
    $\bar{1}\bar{1}011$ & \num{382.895} & \num{0.715} & \num{373.699} & \num{0.818}\\
    $10\bar{2}\bar{1}1$ & \num{121.679} & \num{0.735} & \num{-184.601} & \num{0.902}\\
    $\bar{2}\bar{1}001$ & \num{-133.425} & \num{0.694} & \num{-158.288} & \num{0.508}\\
    $0\bar{2}001$ & \num{113.99} & \num{0.207} & \num{-190.003}  & \num{0.578}\\
    $\bar{1}\bar{2}0\bar{1}1$ & \num{-136.447} & \num{0.574} & \num{159.389} & \num{0.458}\\
    $\bar{2}0\bar{1}\bar{1}1$ & \num{-369.143} & \num{0.724} & \num{-466.979} & \num{0.832}\\
    $0\bar{1}\bar{1}\bar{1}1$ & \num{-141.542} & \num{0.303} & \num{-113.552} & \num{0.780}\\
    $10\bar{1}21$ & \num{318.917} & \num{0.539} & \num{-169.542} & \num{0.552}\\
    $\bar{1}0031$ & \num{210.805} & \num{0.057} & \num{-306.916} & \num{0.646}\\
    $1\bar{1}\bar{1}01$ & \num{-114.081} & \num{0.754} & \num{-85.907} & \num{0.629}\\
    $\bar{1}0\bar{1}01$ & \num{-386.157} & \num{0.057} & \num{-522.279} & \num{0.257}\\
    $01\bar{2}01$ & \num{-109.502} & \num{0.578} & \num{-219.957} & \num{0.659}\\
    $00\bar{1}11$ & \num{183.304} & \num{0.964} & \num{259.756} & \num{0.123} \\
    $\bar{2}0021$ & \num{-468.906} & \num{0.624} & \num{-475.208} & \num{0.718}\\
    $\bar{3}0011$ & \num{-254.551} & \num{0.397} & \num{-212.074} & \num{0.505}\\
    $1\bar{2}011$ & \num{-368.925} & \num{0.519} & \num{412.177} & \num{0.618}\\
    $\bar{4}0001$ & \num{433.004} & \num{0.922} & \num{118.52} & \num{0.71}\\
    $20\bar{2}01$ & \num{125.597} & \num{0.893} & \num{159.223} & \num{0.718}\\
    $0\bar{1}021$ & \num{-126.820} &  \num{0.682} & \num{-102.405} & \num{0.810}\\
    $00\bar{2}\bar{2}1$ & \multicolumn{2}{c}{---} & \num{255.194} & \num{0.921} \\
    $00041$ & \num{-321.194} & \num{0.305} & \multicolumn{2}{c}{---} \\
    \end{tabular}
    \sisetup{round-mode=none}
\end{ruledtabular}
\end{table}


For the \ch{Th:LiCAF} system, we identify one dominant charge compensation pathway $\ch{Th_{Ca}^{..}} + \ch{2 v_{Li}^'}$ in fluoride saturated environments, and two alternative schemes $\ch{Th_{Ca}^{..}} + \ch{Li_{Al}^{''}}$, and $\ch{Th_{Al}^.} + \ch{v_{Li}^'}$ that emerge in a fluoride-deficient environment. In \ch{Th:LiSAF}, the dominant scheme is $\ch{Th_{Sr}^{..} + 2 F_i^{''}}$ in fluoride saturated environments with other schemes, such as $\ch{Th_{Sr}^{..}} + \ch{v_{Li}^{'}} + \ch{F_i^'}$ and $\ch{Th_{Sr}^{..}} + \ch{2 v_{Li}^'}$ also appearing at lower energies, with a larger temperature dependence. In fluoride deficiency, the $\ch{Th_{Al}^.} + \ch{v_{Li}^'}$ scheme seems to emerge with $\ch{Th_{Sr}^{..}} + \ch{Li_{Al}^{''}}$, $\ch{Th_{Al}^.} + \ch{F_i^'}$, and $\ch{Th_{Sr}^{..}} + \ch{2 F_i^'}$ having a similar formation energy at lower temperatures.

\section{Discussion}

Contrary to previous findings in the canonical ensemble~\cite{BandGapCalculPimon2022, LaserExcitatioElwell2024}, our investigation reveals that $\ch{Th_{Ca}^{..}} + \ch{v_{Li}^'}$ and $\ch{Th_{Sr}^{..}} + \ch{2 F_i^'}$ are not universally lowest-energy compensation schemes, but rather in specific conditions --- fluoride-deficient environments for \ch{Th:LiCAF} and fluoride-saturated envioronments for \ch{Th:LiSAF}. We further find that thorium substitution of Ca or Sr appears to be energetically favored in fluoride-saturated conditions.

The low defect formation energy of \ch{Th:LiSAF} indicates that this material can accommodate a higher concentration of thorium defects compared to \ch{Th:LiCAF}. The equilibrium defect concentration at a given temperature in the dilute doping limit is $c(T) = g e^{-G^{f}/{k_{B} T}}$, where $g$ is a configurational degeneracy factor~\cite{FirstPrincipleFreyso2014}. $g$ depends on the defect symmetry and is generally distinct for each structure. For example, the lowest energy structures of \ch{Th:LiCAF} and \ch{Th:LiSAF} in fluoride-deficient environments, $\ch{Th_{Al}^{.}} + \ch{v_{Li}^'}$, show $g = 3$.

Taking the ratio of defect concentrations for these favored compensation schemes at the melting temperatures of LiCAF and LiSAF we find $c^{\ch{Th:LiCAF}}_{\bar{1}0\bar{1}01}(\SI{1082}{\kelvin})/c^{\ch{Th:LiSAF}}_{\bar{1}0\bar{1}01} (\SI{1032}{\kelvin}) \approx 10^{-8}$, suggesting a remarkably larger maximal thorium concentration in \ch{Th:LiSAF}. However, it is essential to note that these values are subject to limitations, including kinetic barriers and systematic errors in DFT calculations. As our approach neglects the impact of ionic mobility, the defect concentration of compensation schemes relying on small-ion diffusion could be underestimated when considering the defect formation energy alone. In addition, as shown in Appendix~\ref{sec:errors}, DFT is prone to inaccuracies exacerbated by the exponential term in $c(T)$. Also note that the dilute doping approximation is invalid for our results of the \ch{Th:LiSAF} system at elevated temperatures due to the low defect formation energy of the favored scheme. While these limitations suggest that the calculated figure might not be exact, the relative advantage of \ch{Th:LiSAF} in accommodating thorium atoms remains a valid conclusion of our results.


An open question remains regarding the adequacy of the compound selection in accurately restricting the chemical potentials. It is likely that other compounds not yet included in the database, or structures that have not been experimentally characterized, will impose further bounds on the chemical potentials. However, we wish to highlight the energy differences involved. Given the relatively small energy difference between the same compensation scheme in the two experimental conditions ($\approx$ 0.3 eV) and the range of calculated defect formation energies (exceeding 1 eV), our analysis provides insights for understanding the potential charge compensation schemes for low-energy defects in Th:LiSAF, despite the limitation of an insufficiently explored phase space.


The EFG values presented in Table~\ref{tab:EFG} enable distinguishing between potential defect structures based on excitation wavelength when using narrow-band laser sources. However, it is worth noting that our analysis focuses solely on the lowest energy structure per compensation scheme and omits configurations with comparable low energies.

An apparent difference from EFG values on thorium reported for \ch{Th:CaF_2}~\cite{Sup229SupTDessov2014} is the observed greater asymmetry values in \ch{Th:LiCAF} and \ch{Th:LiSAF}, which can be attributed to the lower symmetry characteristic of these trigonal materials compared to cubic \ch{CaF_2}.

In conclusion, our DFT calculations reveal that \ch{Th:LiSAF} exhibits lower defect formation energies than \ch{Th:LiCAF} in the grand canonical ensemble, enabling higher doping concentrations. Notably, charge compensation schemes depend on fluorine saturation or deficiency. Our findings align with previous studies for specific conditions: fluorine saturation in \ch{Th:LiSAF} and fluorine deficiency in \ch{Th:LiCAF}~\cite{LaserExcitatioElwell2024, BandGapCalculPimon2022}. We also report EFG values for the lowest energy structure of each charge compensation mechanism.

Our simulations show that due to the numerous possibilities of charge compensation in thorium-doped LiCAF and LiSAF, growing a specific defect structure with desired properties preferentially may be possible by tuning experimental conditions. The selective promotion of defect structures that enhance the electronic bridge effect, where an electron couples to the nucleus, increasing its (de-)excitation rate, is particularly intriguing, as it could lead to significant improvements in the performance of solid-state nuclear clocks~\cite{NuclearExcitatNicker2020, DrivenElectronNicker2021}. In a companion publication~\cite{tobias}, we investigate this concept by comparing electronic bridge transition rates for distinct low-energy charge compensation schemes.

\begin{acknowledgments}
This research was supported by the Austrian Science Fund (FWF) [grant DOI:10.55776/F1004] (COMB.AT) together with the Deutsche Forschungsgemeinschaft (DFG, German Science Foundation) (PA 2508/5-1).
A.P. gratefully acknowledges support from the DFG in the framework of the Heisenberg Program (PA 2508/3-1).
This work has been funded by the European Research Council (ERC) under the European Union’s Horizon 2020 and Horizon Europe research and innovation programme (Grant Agreement No. 856415 and No. 101087184).
The computational results have been achieved using the Austrian Scientific Computing (ASC) infrastructure.
\end{acknowledgments}


\appendix

\section{Defect Formation Energy}
\label{sec:DFE}

\begin{figure}
    \includegraphics{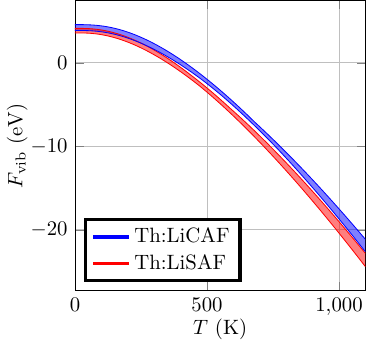}
    \caption{\label{fig:HFE} Ranges of vibrational Helmholtz free energies for the lowest energy structures of each charge compensation scheme.}
\end{figure}

\begin{figure}
    \centering
    \includegraphics{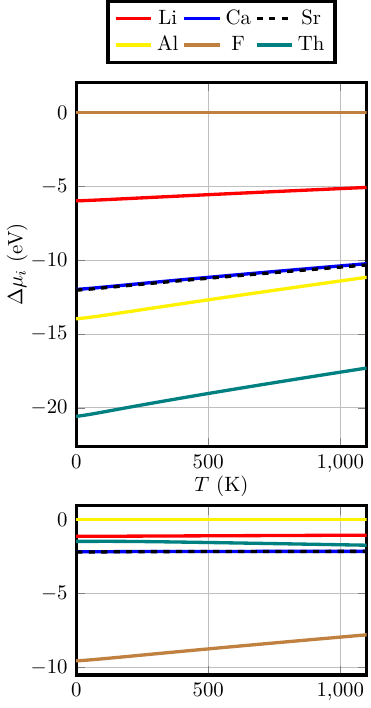}
    \caption{\label{fig:mu} Tuned chemical potentials, free from the pseudopotential reference $\Delta \mu_{i} = \mu_{i} - \mu_{i}[\text{bulk}]$. Top: $\mu_{\ch{F}}$ maximized. Bottom: $\mu_{\ch{F}}$ minimized.}
\end{figure}

This section provides an overview of our approach to calculating defect formation energies. Table~\ref{tab:Uel} presents a summary of calculated electronic energies $U_{\text{el}}$ and the correction terms $E_{\text{corr}}$ for the lowest energy structural configuration per compensation scheme. Fig.~\ref{fig:HFE} displays the corresponding vibrational Helmholtz free energies $F_{\text{vib}}$, which exhibit comparable values across the entire temperature range for all structural configurations. Fig.~\ref{fig:mu} depicts the chemical potentials under the two tuning scenarios. $\mu_{\ch{F}}$ attains its highest possible value when maximized. In contrast, its minimization leads to $\mu_{\ch{Al}}$ reaching its maximum value, accompanied by large values for the $\mu_{i}$ of the other cations.

\begin{table}
\caption{\label{tab:Uel} Electronic energies $U_{\text{el}}[X_{2 \times 2 \times 1}]$ and size correction terms $E_{\text{corr}}$ for the lowest energy structure of each compensation scheme.}
\begin{ruledtabular}
    \sisetup{round-mode=places, round-precision=3}
    \begin{tabular}{c r r r r}
    & \multicolumn{2}{c}{\ch{LiCAF}} & \multicolumn{2}{c}{\ch{LiSAF}} \\
    comp. & $U_{\text{el}}$ (eV) & $E_{\text{corr}}$ (eV) & $U_{\text{el}}$ (eV) & $E_{\text{corr}}$ (eV) \\ \hline
    bulk &  \num{-460.48546044} & & \num{-456.78536472} & \\
    $\bar{1}\bar{1}011$ & \num{-459.588516} & \num{-0.077789} & \num{-456.969682} & \num{-0.789994} \\
    $10\bar{2}\bar{1}1$ & \num{-456.965034} & \num{+1.426208} & \num{-453.635800} & \num{+1.140500} \\
    $\bar{2}\bar{1}001$ & \num{-448.427833} & \num{-0.094332} & \num{-445.219841} & \num{-0.862108} \\ 
    $0\bar{2}001$ & \num{-448.908418} & \num{+0.203692} & \num{-445.948738} & \num{-0.451552}\\
    $\bar{1}\bar{2}0\bar{1}1$ & \num{-436.572505} & \num{+0.261354} & \num{-433.494450} & \num{+0.219294} \\
    $\bar{2}0\bar{1}\bar{1}1$ & \num{-446.400518} & \num{+0.123640} & \num{-443.032069} & \num{-0.476746} \\
    $0\bar{1}\bar{1}\bar{1}1$ & \num{-447.397591} & \num{+0.051387} & \num{-443.894751} & \num{-1.012326} \\
    $10\bar{1}21$ & \num{-479.699583} & \num{-0.217646} & \num{-477.347868} & \num{-0.862627} \\
    $\bar{1}0031$ & \num{-477.996627} & \num{-0.481513} & \num{-476.618469} & \num{-1.686442} \\
    $1\bar{1}\bar{1}01$ & \num{-459.479093} & \num{-0.012497} & \num{-456.497246} & \num{-0.729666} \\
    $\bar{1}0\bar{1}01$ & \num{-458.769698} & \num{-0.283386} & \num{-455.326741} & \num{-0.998958} \\
    $01\bar{2}01$ & \num{-467.960485} & \num{-0.362812} & \num{-463.505018} & \num{-0.898147} \\
    $00\bar{1}11$ & \num{-469.658643} & \num{-0.223978} & \num{-466.841796} & \num{-0.859607} \\
    $\bar{2}0021$ & \num{-467.095884} & \num{+0.513004} & \num{-465.934642} & \num{-0.704817} \\
    $\bar{3}0011$ & \num{-456.595992} & \num{+0.311928} & \num{-453.996442} & \num{+0.204308} \\
    $1\bar{2}011$ & \num{-459.456659} & \num{-0.150109} & \num{-457.662931} & \num{-0.540491} \\
    $\bar{4}0001$ & \num{-445.327762} & \num{-0.231130} & \num{-441.295738} & \num{-0.782057} \\
    $20\bar{2}01$ & \num{-468.097537} & \num{+0.234121} & \num{-465.151050} & \num{-0.169001} \\
    $0\bar{1}021$ & \num{-470.683781} & \num{-0.284090} & \num{-468.583531} & \num{-0.964794} \\
    $00041$ & \num{-488.676618} & \num{+0.135413} & \multicolumn{2}{c}{---} \\ 
    $00\bar{2}\bar{2}1$ & \multicolumn{2}{c}{---} & \num{-442.137026} & \num{+2.006222} \\
    \end{tabular}
    \sisetup{round-mode=none}
\end{ruledtabular}
\end{table}

\section{Chemical Potentials}
\label{sec:chempot}

This section provides additional information on the tuning procedure for the chemical potentials. Table~\ref{tab:Compounds} lists the equilibrium lattice parameters of the standard conventional unit cell for all considered compounds. The table also includes the electronic energy $U_{\text{el}}$ and the pressure-volume term.

\begin{table*}
\caption{\label{tab:Compounds} Optimized lattice parameters, energy per formula unit $U_{\text{el}}$ and pressure-volume term for the selected compounds.}
\begin{ruledtabular}
    \sisetup{round-mode=places, round-precision=2}
    \begin{tabular}{l l r r r r r r}
    Formula & Lattice & $a$ (\si{\angstrom}) & $b$ (\si{\angstrom}) & $c$ (\si{\angstrom}) & $\beta$ & $U_{el}$ (eV) & $pV$ (meV) \\ \hline
    \ch{Li} & cubic & \num{3.434176} & & & & \num{-3.68904487} & \num{0.01280689}\\
    \ch{Ca} & cubic & \num{5.522361} & & & & \num{-6.85328787} & \num{0.02662690} \\
    \ch{Sr} & cubic & \num{6.033265} & & & & \num{-6.35746254} & \num{0.03472191} \\
    \ch{Al} & cubic & \num{4.038930} & & & & \num{-3.75130861} & \num{0.01041706} \\
    \ch{F_2} & gas & & & & & \num{-3.7071223443} & $k_{B} T$ \\
    \ch{Th} & cubic & \num{5.053236} & & & & \num{-7.44470584} & \num{0.02040117} \\
    \ch{LiF} & cubic & \num{4.075260293374} & & & & \num{-11.42907659} & \num{0.01070070} \\
    \ch{CaF_2} & cubic & \num{5.499503780493401} & & & & \num{-22.32263302} & \num{0.02629764} \\
    \ch{SrF_2} & cubic & \num{5.855946587171999} & & & & \num{-21.84494208} & \num{0.03174955} \\
    \ch{AlF_3} & rhombohedral & \num{5.112096845273042} & & & \num{57.90818821798304} & \num{-23.415477825} & \num{0.02843754} \\
    \ch{ThF_4} & monoclinic & \num{13.15670681162} & \num{11.101085591299} & \num{8.602337685047397} & \num{126.31345128020043} & \num{-35.49318265} & \num{0.05335506} \\
    \ch{LiCaAlF_6} & hexagonal & \num{5.0831786235745} & & \num{9.749555} & & \num{-57.560682555} & \num{0.06898623} \\
    \ch{LiSrAlF_6} & hexagonal & \num{5.1663851862269} & & \num{10.374684768} & & \num{-57.09817059} & \num{0.075832495} \\
    \ch{Li_3AlF_6} & monoclinic & \num{14.456764505311} & \num{8.530656} & \num{10.048531000000086} & \num{94.10628710481696} & \num{-58.109531903333334} & \num{0.06514256} \\
    \ch{LiThF_5} & tetragonal & \num{15.236702} & & \num{6.711746} & & \num{-46.97588994375} & \num{0.06158908125} \\
    \ch{Li_3ThF_7} & tetragonal & \num{6.946215} & & \num{12.264524} & &  \num{-67.56082308} & \num{0.093560685} \\
    \ch{CaAlF5} & monoclinic & \num{8.8194414911912} & \num{6.406976} & \num{7.439763000000017} & \num{115.0334364737703} & \num{-45.860649195} & \num{0.06022225} \\
    \ch{Ca_2AlF_7} & orthorhombic & \num{7.0427499435876} & \num{7.7186092316817} & \num{9.601444557037} & & \num{-68.1447910425} & \num{0.0825209375} \\
    \ch{SrAlF_5} & tetragonal & \num{20.20454406} & & \num{14.54130617} & & \num{-45.427434588125} & \num{0.0586580740625} \\
    \ch{Sr_5Al_2F_16} & orthorhombic & \num{7.588576} & \num{12.613898} & \num{14.304355}  &  & \num{-156.383168275} & \num{0.43296634} \\
    \end{tabular}
    \sisetup{round-mode=none}
\end{ruledtabular}
\end{table*}

\begin{figure}
\includegraphics{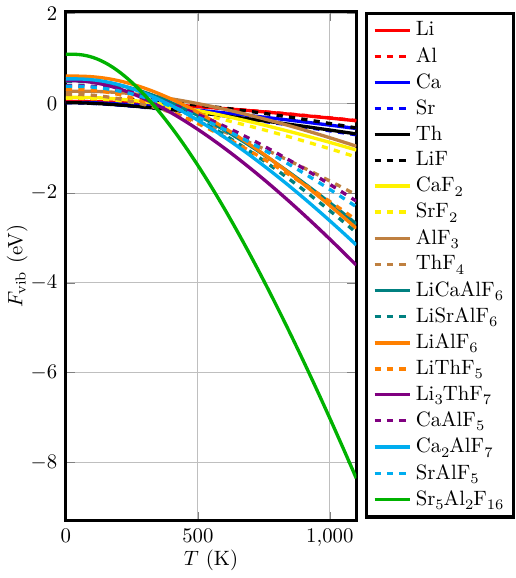}
\caption{\label{fig:Fvib} Vibrational Helmholtz free energy per formula unit for the compounds considered in this study.}
\end{figure}

\sisetup{round-mode=places, round-precision=3}
We utilize the phonopy code for solids to determine the vibrational contributions to the Gibbs free energy (see Fig.~\ref{fig:Fvib}) and employ standard thermodynamic expressions for the gaseous \ch{F_2} phase to calculate the corresponding Gibbs free energy, where the total entropy is a sum of the three components $S = S_{\text{trans}} + S_{\text{rot}} + S_{\text{vib}}$. We derive the translational contribution using the Sackur-Tetrode equation $S_{\text{trans}} = k_B \left(\frac{5}{2} + \ln \frac{k_B T}{p V_Q}\right) $, and the quantum volume $V_Q = \left( \frac{2 \pi \hbar^2}{2m_F k_B T} \right)^{\frac{3}{2}}$. The rotational term is the solution to the quantum rigid rotor $S_{\text{rot}} = k_B\log \frac{2B_0}{k_B T}$, where $B_0 = \frac{\hbar^2}{m_F r_{\ch{F_{2}}}^2}$, and $r_{\ch{F_{2}}}$ is the equilibrium bond length of the \ch{F_2} molecule. For the vibrational part, we use the solution of the quantum harmonic oscillator  $S_{\text{vib}} = k_B\log \left(1-e^{-\frac{\hbar\omega}{k_B T}} \right)$.

\begin{figure}
\includegraphics{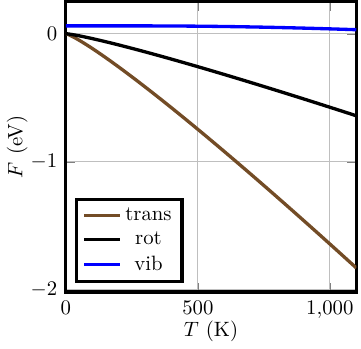}
\caption{\label{fig:F2} Translational, rotational and vibrational contributions to the \ch{F_2} Helmholtz free energy.}
\end{figure}

\begin{figure}
\includegraphics{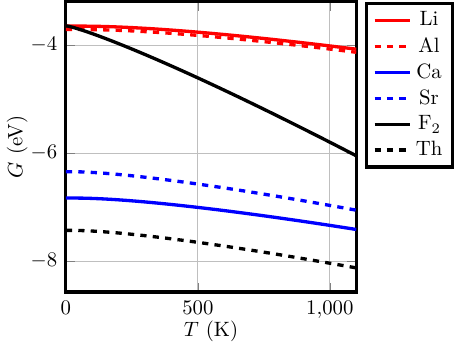}
\caption{\label{fig:mupp} Gibbs free energy of the elements $G_i[\text{metal}] = \mu_{i}$ and $\frac{1}{2}G_{\ch{F_2}}[\text{gas}] = \mu_{\ch{F}}$. Note the inclusion of the pseudopotential reference energies.}
\end{figure}

The fluorine atom mass is our only experimentally determined quantity, which we take as \SI{18.998}{\dalton}~\cite{StandardAtomicProhas2022}. We compute the vibrational frequency and molecular bond length to be $\nu = \SI{29.8817390}{\tera\hertz}$ and $r_{\ch{F_{2}}} = \SI{1.4132725295109996}{\angstrom}$, respectively. In Fig.~\ref{fig:F2} we compare the three contributions to the Helmholtz free energy. Completing our calculation, we add the electronic internal energy $U_{el}$, the vibrational zero-point energy $U_{\text{vib}} = \frac{\hbar \omega}{2}$, and the pressure-volume contribution $pV = k_B T$, to obtain the complete expression for the Gibbs free energy, which we plot in Fig.~\ref{fig:mupp} together with that of the elementary allotropes of the solids in this study.
\sisetup{round-mode=none}

\section{Systematic Errors}
\label{sec:errors}

\begin{figure*}
\includegraphics{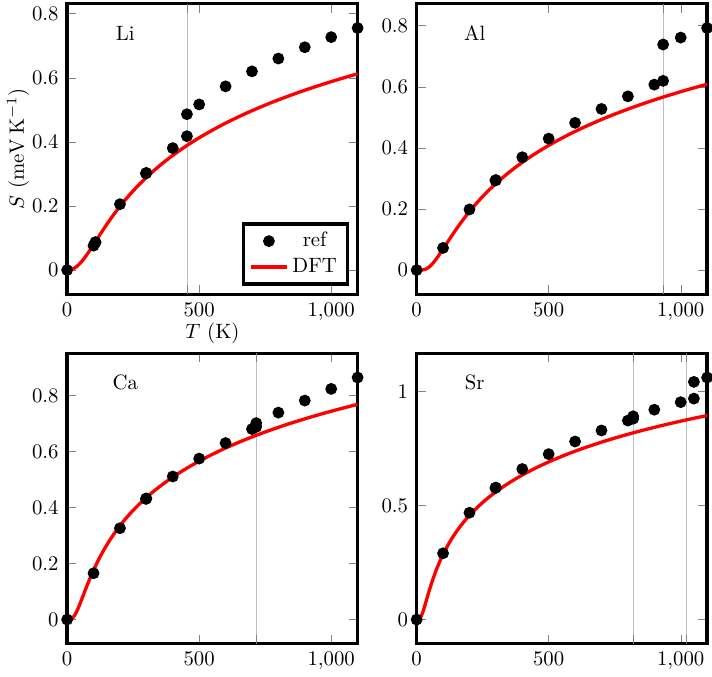}
\caption{\label{fig:ssol} Comparison of calculated entropy to experimental reference values~\cite{nistjanaf}. Highlighted features include phase transitions between crystalline and liquid phases for Li and Al, as well as transitions between $\alpha$ and $\beta$ phases for Ca and Sr.  Additionally, the temperature range of interest shows a crystal-liquid transition in Sr metal occurring close to its upper limit.}
\end{figure*}

\begin{figure}
\includegraphics{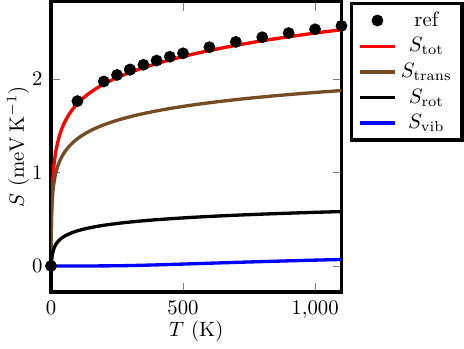}
\caption{\label{fig:sF2} Comparison of the calculated translational, rotational, and vibrational contributions to the entropy of \ch{F_2} and the experimental reference~\cite{nistjanaf}.}

\end{figure}

\begin{figure*}
\includegraphics{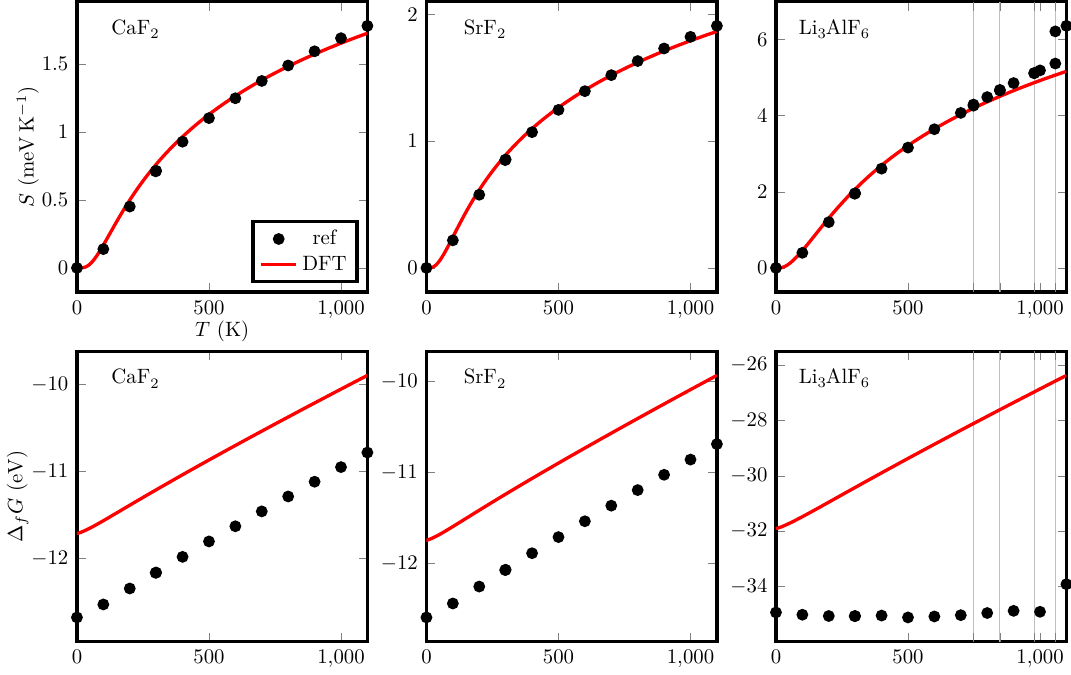}
\caption{\label{fig:sdg} Entropy comparison of the compounds \ch{CaF_2}, \ch{SrF_2} and \ch{Li_3AlF_6} between calculated values in this study, $S^{\text{vib}}$, and experimental reference data~\cite{nistjanaf}. \ch{Li_3AlF_6} undergoes several phase transitions: $\beta \rightarrow \gamma$ at \SI{748}{\kelvin}, $\gamma \rightarrow \delta$ at \SI{848}{\kelvin}, $\delta \rightarrow \varepsilon$ at \SI{978}{\kelvin}, and $\varepsilon \rightarrow \text{liquid}$ at \SI{1058}{\kelvin}.}    
\end{figure*}

This section examines sources of systematic errors in this study. Fig.~\ref{fig:ssol} compares calculated and experimental entropies for elementary solid phases, where the good agreement indicates that neglecting electronic entropy has a minor impact. Notable but small discrepancies arise when materials transition from solids to liquids. Fig.~\ref{fig:sF2} makes a similar comparison for the gaseous \ch{F_2} phase, with generally excellent agreement. In Fig.~\ref{fig:sdg}, we investigate the entropy of selected ionic compounds, and also include a comparison of the Gibbs free energy of formation $\Delta_{f} G$. While the entropy is in line with experiments, we observe well-known systematic deviations in $\Delta_{f} G$~\cite{TheOpenQuantuKirkli2015}, amounting to an overestimation of approximately \SI{10}{\percent} compared to reference values~\cite{nistjanaf}. The error is even more substantial for solids like \ch{Li_3AlF_6}, which exhibit complex behavior due to multiple phase transitions. 

When calculating vibrational properties, we use the mass of thorium-232 instead of the relevant thorium isotope. We evaluate the impact of this approximation by comparing the vibrational free energy of \ch{ThF_4} for both masses. The difference is negligible, with maximum deviations below \SI{2}{\milli \electronvolt} per formula unit at \SI{1100}{\kelvin}.

\bibliography{refs}

\end{document}